# Need to Categorize: A comparative look at the Categories of Universal Decimal classification system and Wikipedia

Almila Akdag Salah[1], Cheng Gao, Krzysztof Suchecki, Andrea Scharnhorst
[1]VKS/KNAW, Cruquiusweg 31, 1019 AT, Amsterdam, the Netherlands.
E-mail: <alelma@ucla.edu>

**Abstract**

This study analyzes the differences between the category structure of the Universal Decimal Classification (UDC) system (which is one of the widely used library classification systems in Europe) and Wikipedia. In particular, we compare the emerging structure of category-links to the structure of classes in the UDC. With this comparison we would like to scrutinize the question of how do knowledge maps of the same domain differ when they are created socially (i.e. Wikipedia) as opposed to when they are created formally (UDC) using classification theory. As a case study, we focus on the category of "Arts".

In modern times, the fast expansion of human knowledge makes categories a necessity in managing and accessing produced knowledge. The science of 'knowledge orders', i.e. taxonomies, classifications, etc., is born out of this need. However today, with all the tools the information society has to offer, taxonomies have a powerful opponent: folksonomies.

Folksonomies are an outcome of the phenomenon of collective writing, and collaborative tagging. Wikipedia is one favorite object for studying such behavior. For a long time, Wikipedia relied only on search engines for information retrieval, and its users browsed the content by following simple links (called page-links) between Wikipedia articles. Only in 2004, after four years of its publication, Wikipedia introduced the concept of categories for the use of its authors. However, what the Wikipedians did by assigning categories to articles, and linking categories to each other, is closer to folksonomies then taxonomies.

Traditionally, classification of knowledge is a task handled by experts, resulting into a designed system of organization. In contrast to this, the category system in Wikipedia is atypical, as it was not designed by experts, but created through the initiatives of individual Wikipedia authors. In this study, we compare the category of Arts from Wikipedia to the main table of Art in the Universal Decimal Classification system. Our goal in this comparison is to address how domains mapped by classification theory differ from socially-mapped domains.

## Wikipedia

Wikimedia Foundation generously shares its monthly backups, which is one of the main reasons why Wikipedia has become a research venue in itself. Not only has the Wikipedia data been applied for many NLP research projects, but also Wikipedia itself as a phenomenon has been studied meticulously from various points: its network structure, growth, nature of its collaborative creation, and the controversies this has fueled.

Wikipedia's category structure is one of the topics that did not get much attention among this research frenzy, and besides a few studies, the topic coverage of Wikipedia is not scrutinized to its depths. Holloway et. al. (2007) compared the top categories and the classification structure of Wikipedia 2005 to widely used encyclopedias like Britannica and Encarta [1]. Halavais et. al. evaluated the topical coverage of Wikipedia by randomly choosing articles, manually assigning categories to them, and mapping the distribution of these to the distribution of published books [2]. A more recent study by Kitter et. al. analyzed the growth of categories, and developed an algorithm to semantically map articles through its category links to the 11 top categories chosen by the research team [3].

Our work follows a similar approach with a focus on category pages and their semi-hierarchy. As noted before, the network of categories is not strictly hierarchical, with clearly defined "top" categories and contains many loops. Still, it possesses vague hierarchical order and it is possible (to an extent) to distinguish this order. To analyze the distribution of articles in "top" categories we had to first define what are these "top" categories.

Looking at the situation in January 2008, we have decided to take Category: Main topic classifications as the root of our category structure. This category-page contains all high-level topical categories. It belongs to higher-level categories itself, which offer different ways of displaying the content of Wikipedia, for example a list of all articles in alphabetical order.

The category network was hierarchized starting from our root. All categories belong to a certain "depth", defined as a distance to the root along the category links. All links that did not follow the hierarchy were discarded (like links between categories at the same depth, or links from a small number to a bigger number in depth level). Then, all articles were given weight of 1. The weight was then propagated up the hierarchical structure using fractional assignment, so that an article page with three categories contributed 1/3 weight unit to each of the three categories. The weights were

**Fig. 1.** The distribution of top categories in Wikipedia (outer) and UDC (inner ring).

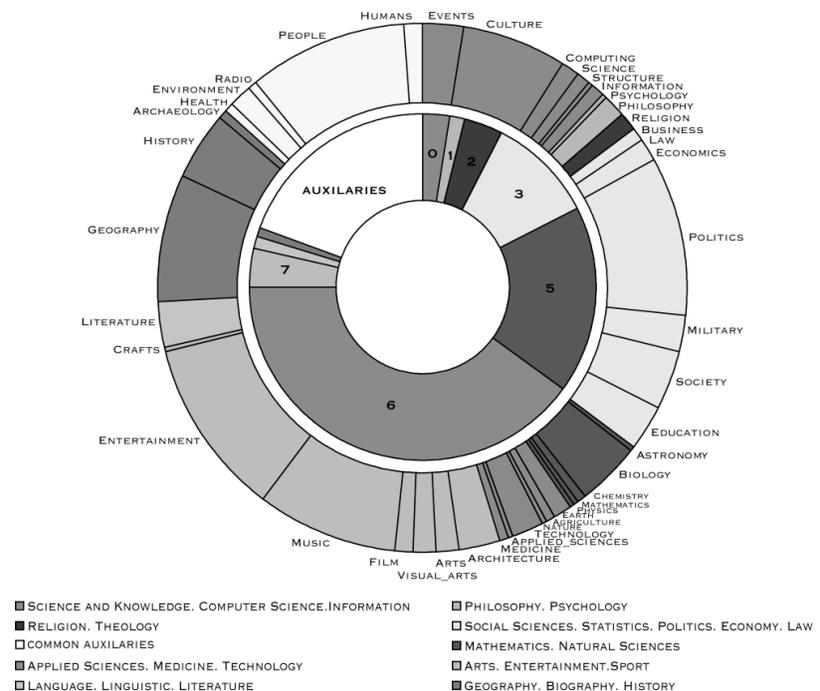

- Science and Knowledge. Computer Science. Information
- Religion. Theology
- Common auxiliaries
- Applied Sciences. Medicine. Technology
- Language. Linguistic. Literature
- Philosophy. Psychology
- Social Sciences. Statistics. Politics. Economy. Law
- Mathematics. Natural Sciences
- Arts. Entertainment. Sport
- Geography. Biography. History

propagated to the level of our "top" categories. Since we used fractional assignment, the sum of the weights equals the total number of articles found in the whole hierarchical network under our root category.

Figure 1 shows the distribution of all category pages to our root categories, which were 43 in total (the outer ring). In contrast to this abundance at the root level, classical classification systems have a defined root. The UDC knows 9 top categories called main tables. Figure 1 depicts these main tables of UDC as well (the inner ring). To ease the comparison, we have mapped Wikipedia categories to UDC tables. This exercise demonstrated the fact that most of the 'top' categories of Wikipedia belong to one of the main tables of UDC at the second level. However, certain categories like People, Humans, Health, Environment do not have a direct equivalent in UDC. Our findings confirm Kitter et. al.'s in that applied sciences and technology are underrepresented in Wikipedia, whereas topics related to popular culture have a high coverage percentage.

## UDC

The foundation of UDC goes back to two Belgian lawyers, Paul Otlet and Henri La Fontaine, who as early as 1895 envisaged a classification system that should be able to organize all existing knowledge [4]. Since the first 1905 edition, the existing 19$^{th}$ century structure of UDC has been filled with overlapping concepts from the 20$^{th}$ century [5]. From 1993 onwards, UDC consortium publishes a digital Master Reference File [MRF] as UDC standards. Our data stems from this master files 2008 edition. In 2008, MRF has reached the record number 68546, and out of this number only about 4 percent, (2601 items) was devoted to the main table 7: Arts, Entertainment, Sports. Figure 2 depicts the distribution of the UDC main table 7 (in light gray) as well as Wikipedia's category Arts (in black) according to their depths.

In Figure 2 we have included a third category which we call 'Arts, combined' (in dark gray), this corresponds to all the top categories in Wikipedia such as Visual Arts, Architecture, Music, Film, Crafts, Entertainment etc. that according to UDC classification would have belonged to Arts. The category tree of UDC does not branch more than to the 9$^{th}$ level, and follows almost a Gaussian distribution. The distribution of Arts combined follows almost a similar curve to UDC's Arts category and has the last subclasses on the level of 10, whereas Wikipedia Arts has a sharp increase on the 5$^{th}$ level, and dies out on the 7$^{th}$ level [6].

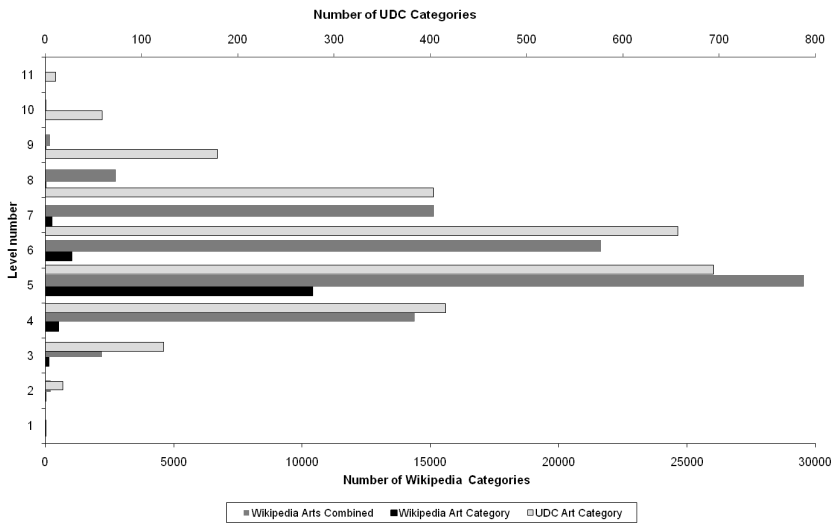

Fig. 2. The sub-category distribution of Wikipedia & UDC in Arts

The Wikipedia category system is not a pure tree graph, but can be rather depicted as an overlay between different tree graphs. However, forced into a tree structure, Wikipedia's category tree for Arts encompasses 2361 category pages. When we include all art-related categories as in Arts-combined, this number increases to 86133. All in all, in Wikipedia more sub-classes are devoted to Arts, more than the whole UDC offers to librarians to classify their collections. But, UDC numbers can be extended, combined and changed when applied to collections. That leads to the question if an analysis of the use of the UDC for art collections in a comparable bottom-up perspective would reveal a greater variety than the MRF shows.

## Conclusion

As reported earlier the topic coverage of Wikipedia falls relatively short on topics devoted to natural and applied sciences, and is more focused on "general" topics of human interest. Nevertheless most of Wikipedia's main topic category terms have a relative clear counterpart in UDC. Moreover, the emphasis on topics of general interest is very much in line with the original envisioned structure of the UDC as an indexing "language" for all types of knowledge. By analyzing and comparing different classification systems we get new insights about the collections as well as their intended audience. By mapping Wikipedia categories into UDC classes we also learn about the differences in their nature. Concerning the category of Arts we found a similar pattern of differentiation looking at the distribution of Wikipedia articles and of UDC numbers across depth levels. The large number of subcategories belonging to Wikipedia category "Arts-combined" forces one to question the use of such a bottom-up designed categorization system. We plan to follow this question by analyzing in detail the subcategories of Arts in both systems.